# Clusters and Maps of Science Journals

# Based on Bi-connected Graphs in the *Journal Citation Reports*

Loet Leydesdorff

Science & Technology Dynamics, University of Amsterdam

Amsterdam School of Communications Research (ASCoR)

Kloveniersburgwal 48, 1012 CX Amsterdam, The Netherlands

loet@leydesdorff.net ; http://www.leydesdorff.net

**Abstract**

The aggregated journal-journal citation matrix derived from the *Journal Citation Reports 2001* can be decomposed into a unique subject classification by using the graph-analytical algorithm of bi-connected components. This technique was recently incorporated in software tools for social network analysis. The matrix can be assessed in terms of its decomposability using articulation points which indicate overlap between the components. The articulation points of this set did not exhibit a next-order network of 'general science' journals. However, the clusters differ in size and in terms of the internal density of their relations. A full classification of the journals is provided in an Appendix. The clusters can also be extracted and mapped for the visualization.

## 1. Introduction

In an article in *Science* entitled "Networks of Scientific Papers," Derek de Solla Price (1965) reported that he had been able to work with the experimental version of the *Science Citation Index* for 1961. In this paper, he noted the possibility of studying the dynamics of journal relations by using the aggregates of their mutual citations. However, it would take until 1975 before the Institute of Scientific Information began to compile this data systematically on a yearly basis (Garfield, 1979).

By now, the *Journal Citation Reports* of the *Science Citation Index* covers some 5500+ journals. The decomposition of this data into journal groupings is urgent for reasons of improving both the consistency in journal collections and the baselines in scientometric evaluations (Studer & Chubin, 1980; Moed *et al*., 1985; Glänzel & Schubert, 2003). Journal sets can be considered as indicators of the intellectual organization of the sciences at the aggregated level (Leydesdorff, 1987). Furthermore, they are reproduced from year to year with considerable stability, and hence they can be used also as indicators of evolutionary change (Leydesdorff, 2002a). However, the problem of how to delineate these sets consistently and over time has been a major concern of the scientometric analysis since its early days (Narin *et al.*, 1972; Carpenter and Narin, 1973; Narin, 1976; Leydesdorff, 2002b).

The data can be considered as a huge matrix of the aggregated citations among the journals. Each cell *ij* of the matrix indicates how often journal *i* cites journal *j* during a given year. Such a large matrix, however, could hitherto not be analyzed easily with software because of computational limitations. It had to be broken down into chunks, or specific algorithms had to be used in order to address local densities contained in the matrix. The choice of one algorithm over another, or delineation in terms of one subset or another, however, implies a selection, and therefore various representations of this data could be entertained. A unique solution seemed impossible given the limitations in hard- and software.

In an early stage, the ISI developed indices like the impact factor of journals (Garfield, 1972; Garfield and Sher, 1963) and the immediacy index (Price, 1970; Cozzens, 1985; Moed, 1989). The question of clustering the data was raised in the 1970s by Narin *et al*. (1972) and by Carpenter and Narin (1973). These authors, however, focused on mapping the hierarchy among the journals (Narin, 1976). A better understanding of stratification in the journal set would enable libraries to rationalize their portfolios given budget constraints (e.g., Hirst, 1978), and it might also provide us with a baseline for comparisons in research evaluations (Studer & Chubin, 1980; Moed *et al.*, 1985).

Given the computational limitations at that time the approaches of the 1970s were based either on using aggregate measures like "total cited" or "total citing" (e.g., divided by the total number of publications in order to compute the citation/publication-ratios of groups of authors or the impact factors of journals) or algorithms that are essentially based on "single-linkage clustering." Single-linkage clustering can be based on sorting the links (that is, cell values) into a list. A list can be indexed by decreasing frequencies using software for database management. Thus, the two-dimensional problem of decomposition of the matrix could be reduced to the one-dimensional problem of handling lists.

For example, co-citation analysis was developed within ISI to map the sciences using single-linkage clustering (Small & Griffith, 1974). The mapping in more than a single dimension, however, requires a two-dimensional visualization technique (Small & Sweeney, 1985; Small *et al.*, 1985; cf. Leydesdorff, 1987; Small, 1999; Chen, 2003). A multi-dimensional approach using the relations among the journals as a network of links and the cited journals as the nodes became more widespread during the 1980s. Doreian & Fararo (1985) used graph analytical techniques for block-modeling the relations, but the focus of these studies was still heavily on discovering the relative standing of journals within the hierarchy (Doreian, 1986).

Tijssen *et al.* (1987) used quasi-correspondence analysis to map the journals as groups on the basis of their deviation from normalized expectations. Clusters of journals could then be made visible. However, the delineations had to be penciled into the pictures and remained based on intuitive or expert-based classification. One advantage of this method was that it allowed for the visualization of both the cited and the citing patterns in a single mapping. Leydesdorff (1986 and 1987) used factor analysis on selective parts of the aggregated journal-journal citation matrix. The factor analysis provides us with clear delineations, but this technique remains limited in terms of the number of variables that can be rotated in each run.

These various authors agreed in using the information contained in the matrix of aggregated journal-journal relations, but they did not pursue the analysis with similar interests. The data contain two types of structures (Burt, 1982), namely: (a) a hierarchical structure that can be studied at the level of the database (e.g., in terms of impact factors) or at the level of specific groups of journals (e.g., *The Lancet* and *The New-England Journal of Medicine* for the medical sciences; the *Journal of the American Chemical Society* within chemistry); and (b) a grouping in terms of disciplines and specialties. The different groupings cannot be expected to entertain hierarchical relations across groups, but certain journals may relate to other groups as a layer of next-order, i.e., multi-disciplinary, journals (e.g. *Science, Nature, PNAS USA*; cf. Glänzel & Schubert, 2003).

The standing of a journal in the hierarchy is a property of the journal in relation to its competitors, but the grouping is a function at the network level. The factor (or cluster) analysis enables us to identify eigenvectors of the network that can be considered as orthogonal dimensions. These dimensions are independent. Ranking and grouping are thus two very different operations. The groupings remain heavily dependent on the specific delineations and sensitive to changes in the journal citation structures over time (Leydesdorff & Cozzens, 1993). Thus, they cannot easily be reproduced for different years; the ranks can be expected to show more stability at the top of the hierarchy, for example, in time-series and trend analysis.

Recent developments in social network analysis and the availability of larger capacities in terms of computer memory management enable us nowadays to make further progress in handling the data contained in the *Journal Citation Reports* both hierarchically and in terms of their structure. First, the entire network can be read into a computer program for the statistical analysis or the visualization because the various programs in the Windows environment can nowadays use all internal memory available (Leydesdorff, 2004; Leydesdorff & Jin, in preparation). Second, the algorithm for bi-connected components was recently incorporated in software tools for social network analysis (Moody & White, 2003). This technique was developed in order to find robust clusters in large data sets (Knaster & Kuratowski, 1921).

## 2. Bi-connected component analysis

A bi-connected component is a maximally connected subgraph in which for every triple of vertices *a*, *v*, and *w* there exists a chain between *v* and *w* which does not include the vertex *a*. In other words, each node in the bi-connected component is linked to at least two other nodes in this cluster. Therefore, the network remains connected after removing any vertex (Mrvar & Batagelj, *n.d.*). This bi-connectedness stabilizes the cluster against changes in the initial selection when producing the database. Thus, the inclusion or exclusion of journals by ISI would not directly affect the large bi-components contained in the network data.

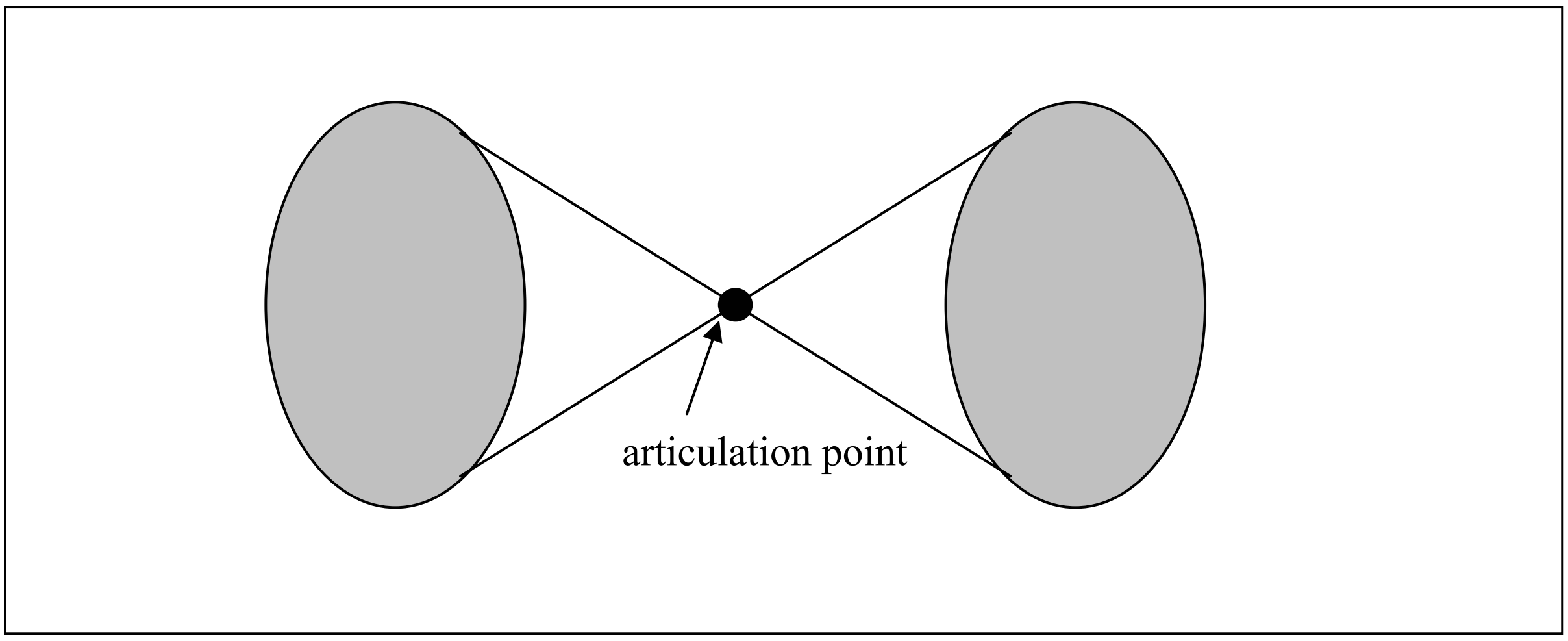

**Figure 1**

Two bi-connected network components with an articulation point

Between two bi-connected components of a network, there can be an overlap. A vertex in the overlap is called an *articulation point* or cut-point. A node in a graph is an articulation point if removal of this node breaks the graph into more than one bi-connected component (Scott, 1991).[1] Articulation points belong to more than a single bi-connected component and can therefore be considered as a next-order in the hierarchy. A visualization program like Pajek enables us to study both bi-connected components and articulation points by selecting partitions from the data file.[2]

## 3. Methods and materials

The *Journal Citation Report 2001* was the most recent one available on CD-Rom at the date of this research (April 2003). It contains citation data for 5,748 journals. Of the 5,748 x 5,748 possible cells only 860,374 (2.6%) are listed with a value.[3] Thus, the citation matrix is very sparse. We know from previous research that the journals occur in dense clusters which indicate specialties with only exceptional links between these structures. The purpose of this study is to decompose the matrix and to study the linking journals as a next-order level of hierarchical integration. To what extent does hierarchical integration operate and counterbalance the ongoing processes of disciplinary differentiation?

The Pearson correlation will be used as similarity measure among the citation patterns. This is based on the pragmatic consideration that this measure is available in the Ucinet software. It has been noted that the Pearson correlation is sensitive to the number of zeros (Ahlgren *et al.*, 2003) while the cosine is not. Thus, in a further improvement one may wish to use the cosine. The difference between the two measures, however, can be expected to be relatively small in terms of the eventual results (Leydesdorff & Zaal, 1988; White, 2003).[4]

---

[1] The definitions provided in this section apply to both directed and undirected graphs.
[2] The program Pajek is freely available for non-commercial use at http://vlado.fmf.uni-lj.si/pub/networks/pajek .
[3] Although 216 cells with the value of one could be counted in the data, this value seems normally truncated in the electronic (CD-Rom) version of the *JCR*. Above the value of one, all cell values seem included in the data since 244,776 cells have a value of two, 129,920 a value of three, etc.

We decided not to normalize the main diagonal values that indicate the journal "self-citations." The self-citations are known to be outlayers and algorithms for the normalization have been proposed (Noma, 1982; Price, 1981). In the first round of the descriptive statistics, however, we have wished to stay as close as possible to the information contained in the data. After the normalization (for example, using the Pearson correlation) the main diagonal is anyhow no longer included in the analysis.

In summary, we use both the aggregated journal-journal citation matrix and the Pearson correlation matrix of the citation patterns in the citing dimension as input to the analysis. The citing dimension is here analyzed first, because active citation can be considered as the running variable that is year-specific, while "being cited" refers to the status of the journal in the database. However, the analysis can be repeated in the other dimension by transposing the citation matrix. We intend to pursue the analysis in this dimension in a follow-up study.

First, we analyze the aggregated citation matrix. Threshold levels can be used in order to observe the clusters more distinctly. Second, we turn to the Pearson correlation matrix. Since we are interested in the clusters and not—in this study—in the distances among the clusters, we focus on the positive correlations. Again, one can vary the threshold level. In a third step we pursue the hierarchical analysis using bi-connected components at a high level of correlation (e.g., $r \geq 0.8$).

[4] One major difference is that the Pearson correlation between two empty columns of zeros is necessarily equal to unity, while the cosine is equal to zero in this case. Thus the use of the Pearson correlation can be expected to generate the artifact of a large bi-connected graph among empty vectors. Indeed, a large cluster of 230 journals which were not processed by ISI, but included in the database, was found in this data.

The networks are read into Pajek for the visualization. Ucinet 6 is able to generate the huge Pearson correlation matrix from this data (5,748 variables),[5] and it allows for exportation of the results in a format that can be imported into Pajek (Borgatti *et al*., 2002). The visualizations are based on two algorithms available for this purpose in the latter program. One algorithm (Kamada & Kawai, 1989) uses the least squares method under the condition of a set of conflicting constraints. It provides clear pictures in relatively small sets (< 100). The other (Fruchterman & Reingold, 1991) is based on the metaphor of a system of interacting particles with repelling forces. This algorithm then minimizes the energy (Owen-Smith *et al*., 2002).

## 4. Results

### *4.1 The data matrix*

Of the 5,748 journals included in the *Journal Citation Reports 2001*, nine are not cited at all and 26 relate to each other in 13 bi-graphs only. The 5,713 remaining journals form one single network which is exhibited in Figure 2. Unfortunately, one is not able to read the labels other than at the margins and surfaces because the points and labels overlap one another in the central regions.

[5] This is exceptional. MS Excel, for example, has a limitation of 256 (= $2^8$) columns.

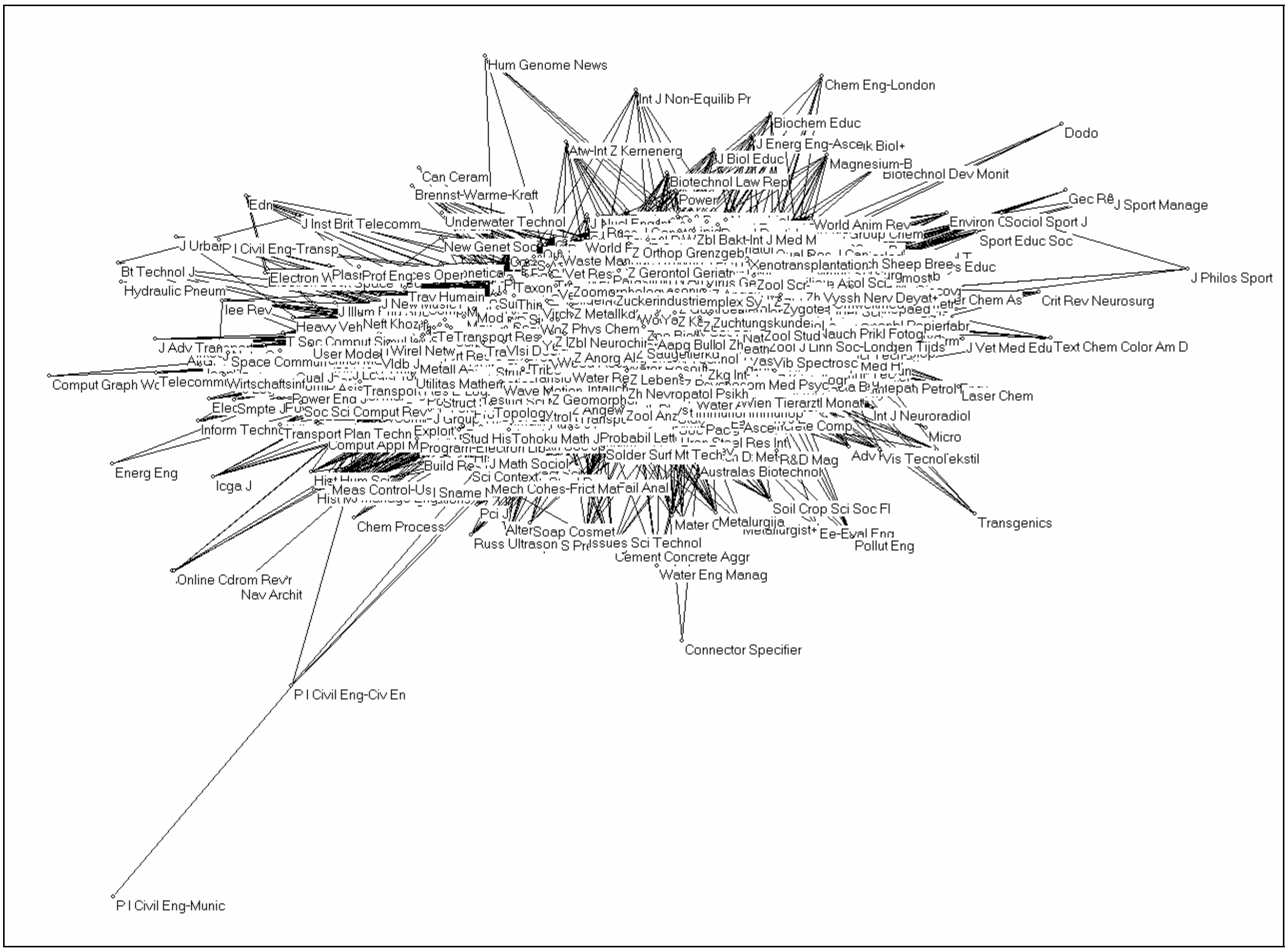

**Figure 2**

Two-dimensional solution of the *JCR 2001* data matrix (5,713 journals) using the Fruchterman-Rheingold algorithm

Figure 3 shows the effect of raising the citation threshold, for example, to 100 citations. In this case only 3,529 (61.4%) journals are still connected to the central network. Note that there are a number of small isolated networks, while some medium-sized networks are only weakly connected at this threshold level. However, we will pursue the analysis of these groups using normalized citation rates.

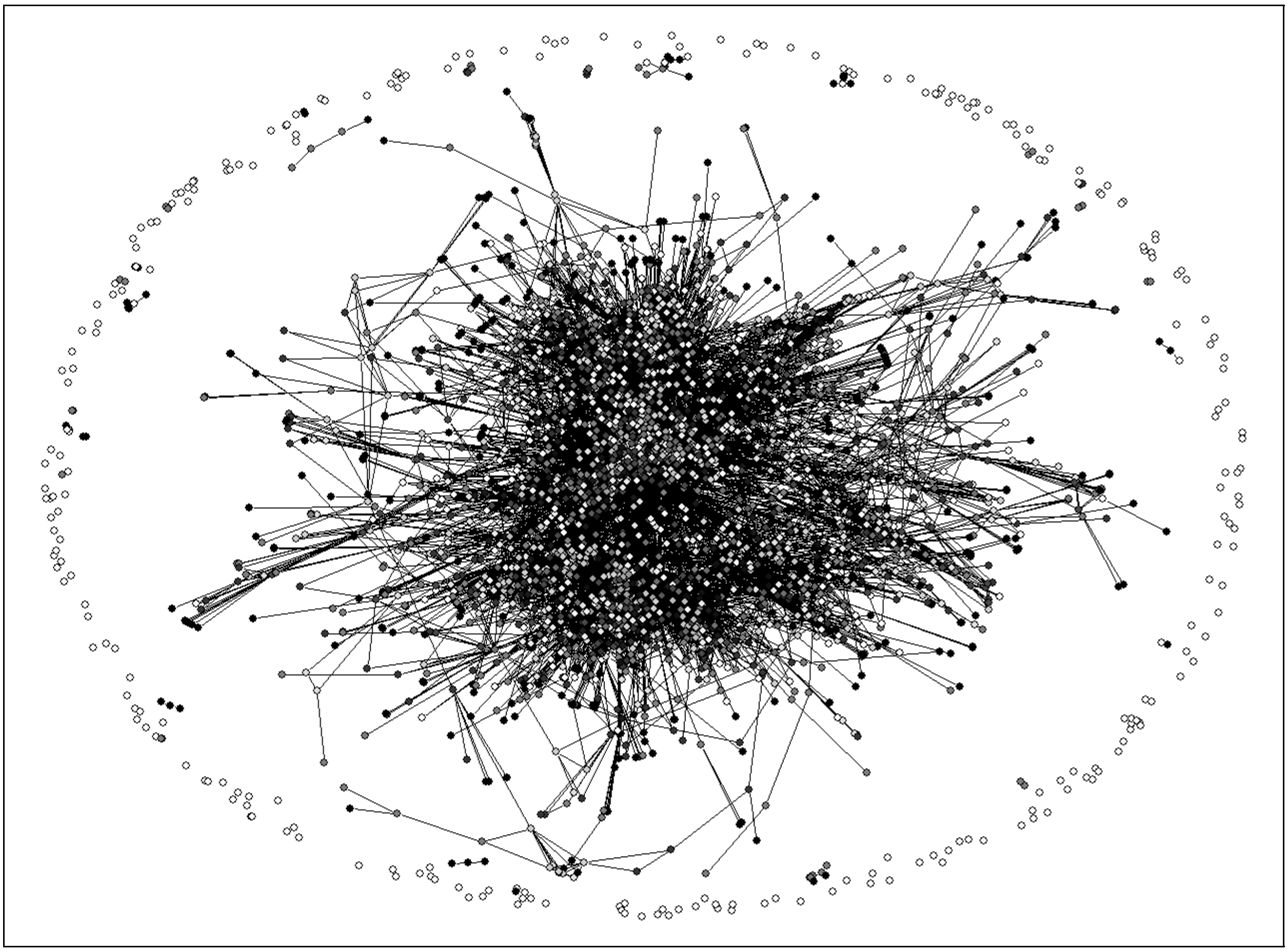

**Figure 3**

Idem after removal of citation relations < 100

*4.2 Normalized citation rates*

The Pearson correlation can vary between -1 and +1. In our case, the negative values indicate dissimilarity. In this study, however, we focus on the groupings in terms of similarity. (The negative values provide us with an indicator of the distances between the groups.) Figure 4 provides the solution using a correlation between citation patterns of ≥ 0.5 as a threshold. First, 230 journals were removed that are not processed by ISI in the "citing" dimension, or to such a low extent (fewer than twelve times) that the Pearson *r* evaluates to unity.[6] This reduces the sample to (5,748 – 230 =) 5,518 journals. 5,405 of these journals (98.0%) are connected to the central network at least at one place with $r \geq 0.5$. Only very few and small groups are disconnected at this level of correlation (Figure 4).

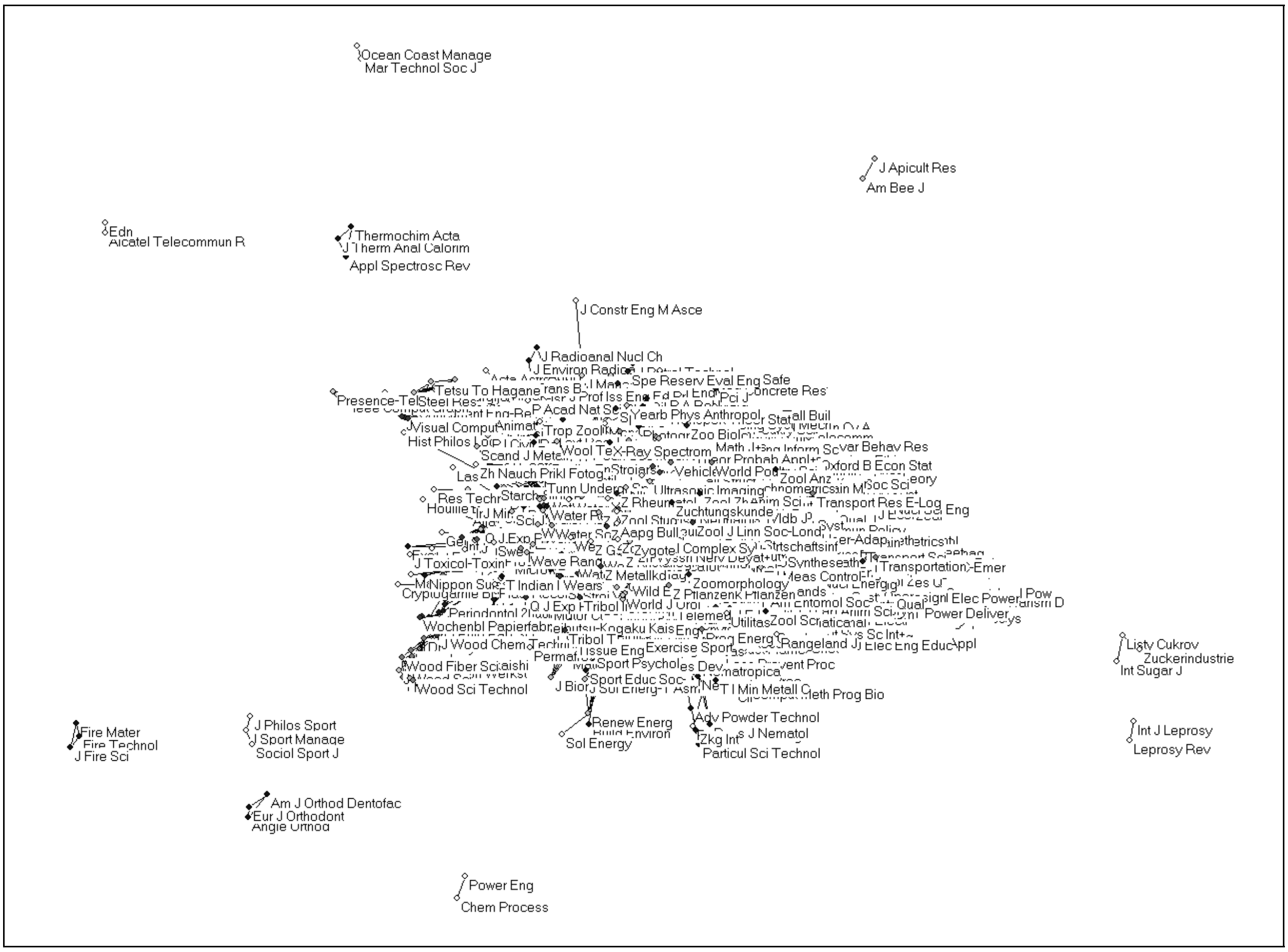

**Figure 4**

5,405 journals from the *JCR 2001* relating to one another in citation patterns with $r \geq 0.5$

In other words, almost all journals correlate with some other journals in their citation patterns to a high degree. Even if the threshold is raised to $r \geq 0.8$, 3,991 (73.8%) remain included in the analysis. An increasing number of components, however, form subgroups which correlate only at the specialty level. These are indicated in Figure 5 in white. However, the main cluster remains a huge network. As we shall see below, this central network contains all the bio-medical journals, but it reaches far beyond this set.

[6] The *Journal Citation Report 2001* of the *Science Citation Index* gives only for 20 journals a total number of citations equal to zero. 226 journals are indicated as citing between zero and ten times, and two more only eleven times. These 230 are excluded from further analysis because the *r* between their citation patterns is larger than 0.99.

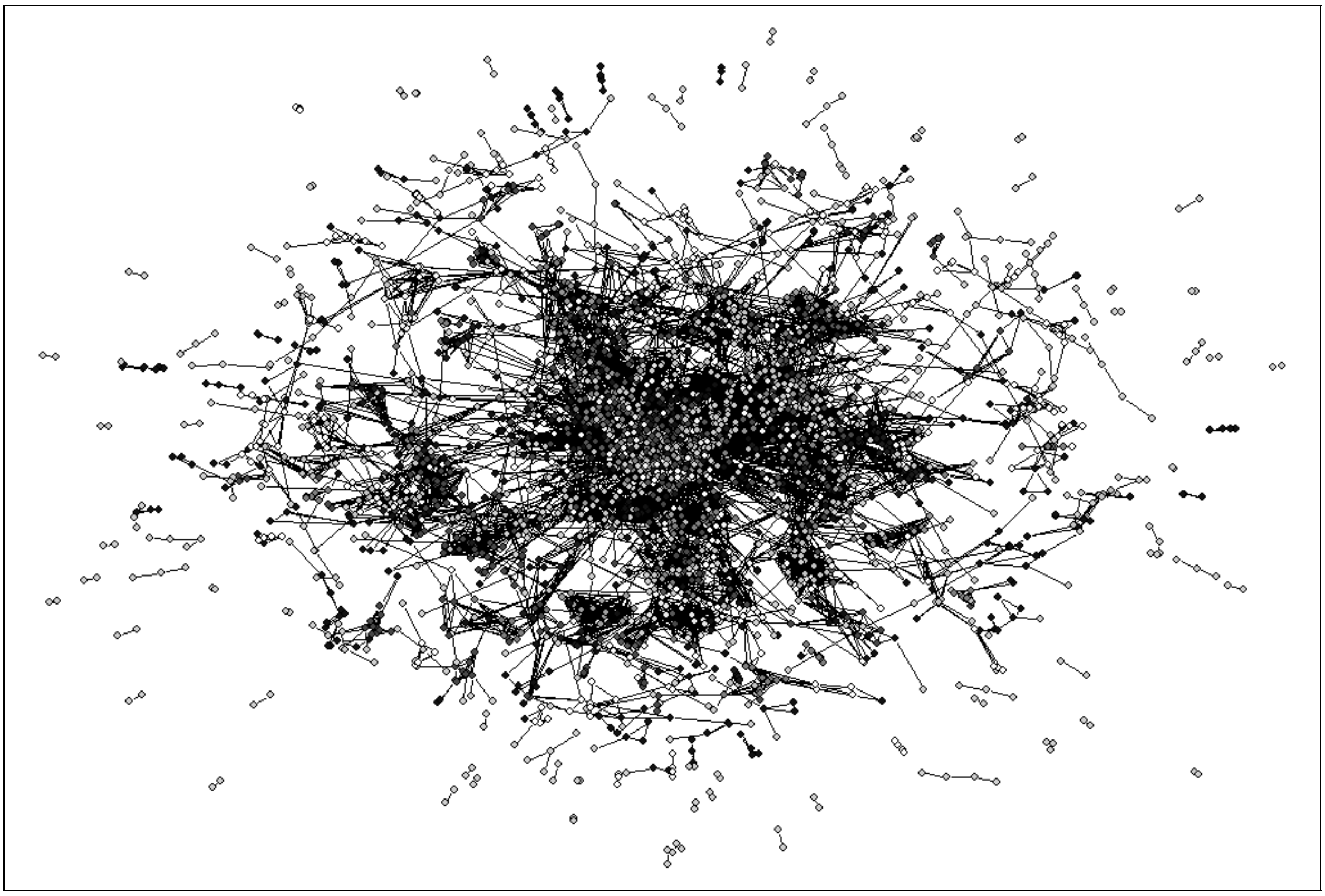

**Figure 5**

3,991 journals from the *JCR 2001* relating to each other's citation patterns at $r \geq 0.8$

I experimented with raising the threshold even further. However, at the level of $r \geq 0.9$ only 2,440 journals (45.1%) are included in the total set. Therefore, we will begin the decomposition in the next section at the level of $r \geq 0.8$, that is, with 3,991 journals included.

Narin *et al*. (1972) argued that most journals are strongly related to the central network at one place or another by a next-order type of journals. Thus, a hierarchy would be formed. Our results from using factor analysis (Leydesdorff & Cozzens, 1993) pointed also in this direction. For example, while "hydrobiology" and "marine biology" are indicated as separate clusters by using factor analysis of the local citation environment of the journal *Limnology and Oceanography*, this journal itself functions as a part of both these groups. In the rotated factor matrix this is visible (in Table I) as interfactorial complexity of the factor loadings of this journal's citation pattern. *Limnology and Oceanography* can thus be considered as a link between these networks.

Rotated Factor Matrix:

| | FACTOR 1 | FACTOR 2 | FACTOR 3 | >>>> |
|---|---|---|---|---|
| *Mar Ecol-Prog Ser* | **.95793** | .13870 | .08140 | |
| *J Exp Mar Biol Ecol* | **.91895** | -.11895 | -.03447 | |
| *Mar Biol* | **.89163** | -.05321 | -.02910 | |
| *Estuar Coast Shelf S* | **.77746** | -.16425 | -.04488 | |
| *J Plankton Res* | **.60994** | .35307 | .44847 | |
| *Deep-Sea Res Pt I* | .09786 | **.91904** | -.18993 | |
| *Deep-Sea Res Pt II* | .20122 | **.91303** | -.06394 | |
| *J Geophys Res* | -.21268 | **.61808** | -.26037 | |
| ***Limnol Oceanogr*** | **.42113** | **.59319** | **.37911** | ◄ |
| *Freshwater Biol* | -.15609 | -.08635 | **.91136** | |
| *Arch Hydrobiol* | -.16927 | -.09394 | **.91092** | |
| *Hydrobiologia* | .31555 | -.02682 | **.71990** | |

**Table I.** Three first factors in the 1% citation environment of *Limnology & Oceanography* 2001. (See Leydesdorff & Cozzens (1993) for details about the methodology.)

In summary, one expects both grouping and ranking in this data. However, normalization using the Pearson correlation did not provide us with a sufficient instrument for observing either the structural dimensions or the hierarchy in the network of relations. For analyzing structure one needs the rotation of the factor analysis, and for analyzing hierarchy one needs a hierarchical clustering algorithm.

### *4.3 Bi-component clustering*

As noted, 3,991 journals are included in the analysis at the level of Pearson's $r \geq 0.8$.[7] When we run the analysis on the 2001 dataset, 222 components are bi-connected at the indicated level. (A bi-connected component by definition consists of three journals at the minimum.) These 222 bi-components include 3,447 of 3,991 journals with 55 articulation points among them. Figure 6 provides a visual representation (using the format of Figure 5) with the articulation points in white. The articulation points are thinly spread over the map. In other words, most bi-components are not related to a higher-level network of coordination. We return to this issue below.

[7] The (5,405 – 3,991 = ) 1,414 journals not related at this level may contribute to the dynamics of journal-journal structures over time, but they do not have an effect on the structure of this data in 2001 (Leydesdorff, 2003).

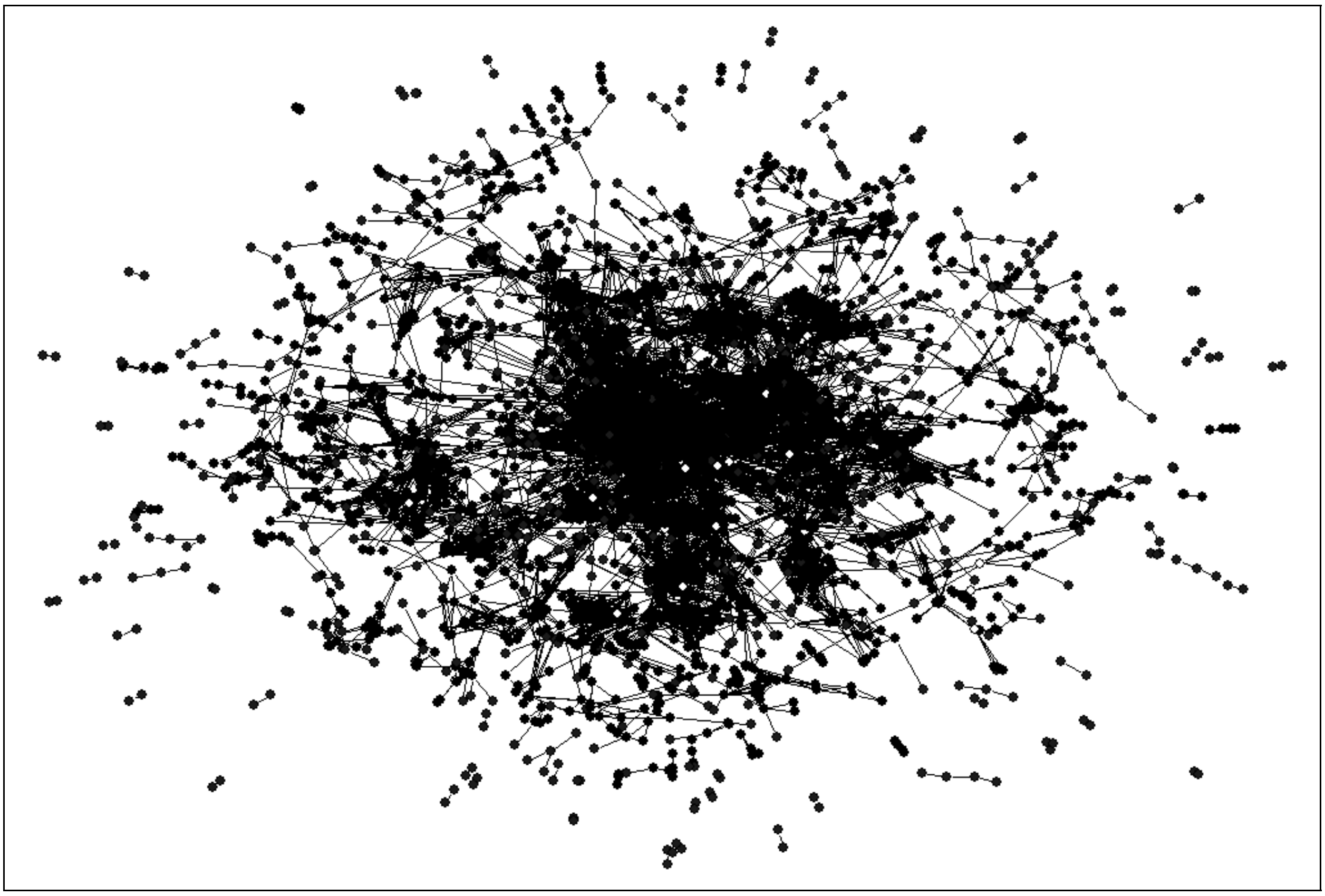

**Figure 6**

The 55 articulation points are specifically colored in white.

Figure 7 first shows the distribution of the bi-connected components in terms of their sizes.

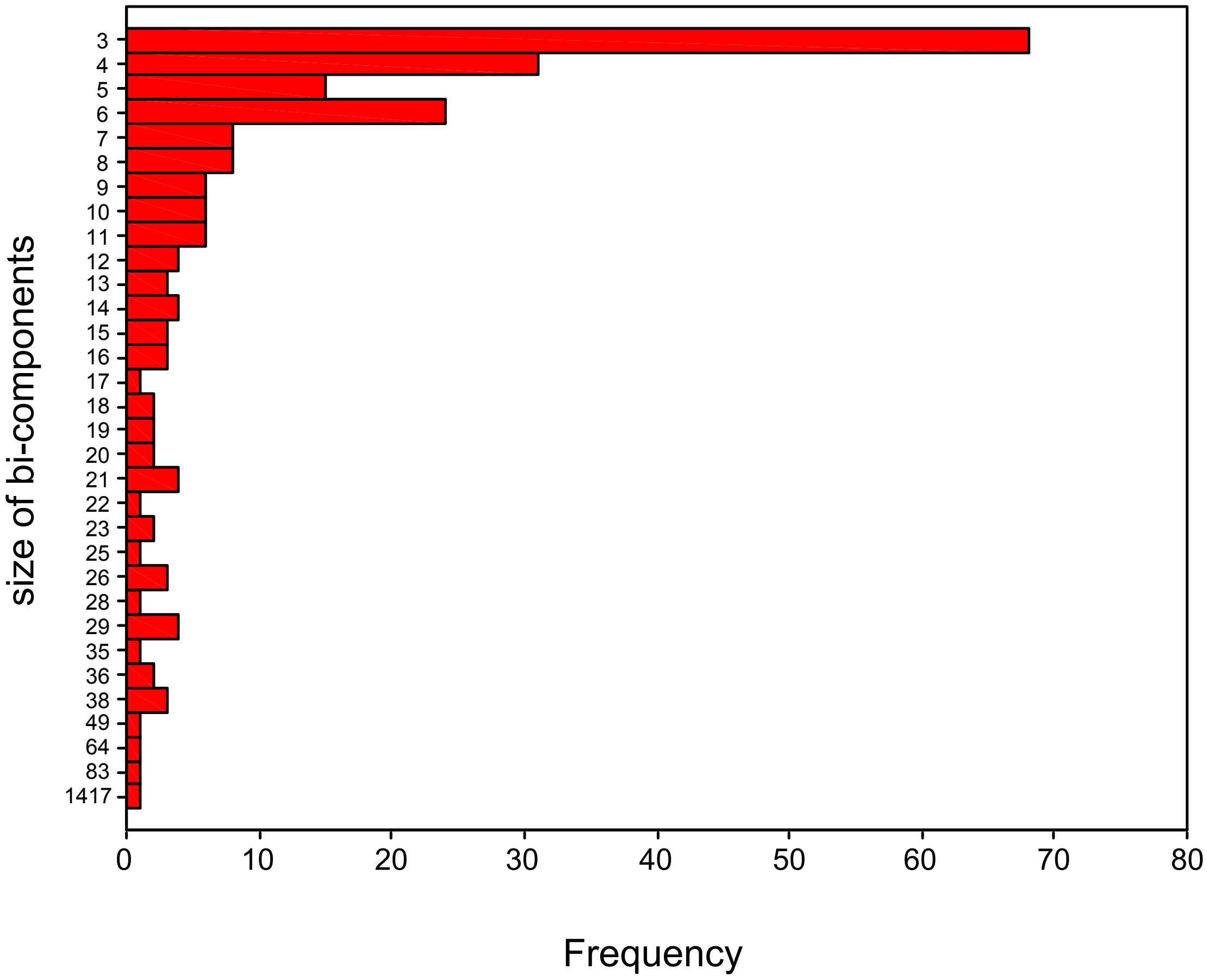

**Figure 7**

Size distribution of 222 bi-connected components contained in the data.

One cluster is disproportionally large, containing 1,417 journals. This cluster will be further analyzed in the next section by raising the threshold to $r \geq 0.9$. Other clusters are smaller and very homogenous. Analogously, if we lower the threshold, for example, by including *r*-values between 0.7 and 0.8, we would merge a large number of the smaller clusters, but the specificity then also decreases.

Let us, for example, focus on the one bi-component with 49 journals. This cluster contains the following titles:

| | | |
|---|---|---|
| *Acta Polym Sin* | *J Polym Eng* | *Plast Rubber Compos* |
| *Adv Polym Sci* | *J Polym Mater* | *Polimery-W* |
| *Adv Polym Tech* | *J Polym Res-Taiwan* | *Polym Advan Technol* |
| *Biomacromolecules* | *J Polym Sci Pol Chem* | *Polym Bull* |
| *Chinese J Polym Sci* | *J Polym Sci Pol Phys* | *Polym Degrad Stabil* |
| *Comput Theor Polym S* | *J Soc Rheol Jpn* | *Polym Eng Sci* |
| *Des Monomers Polym* | *Kobunshi Ronbunshu* | *Polym Int* |
| *Eur Polym J* | *Korea Polym J* | *Polym J* |
| *High Perform Polym* | *Macromol Biosci* | *Polym Polym Compos* |
| *Int J Polym Anal Ch* | *Macromol Chem Phys* | *Polym Test* |
| *Int Polym Proc* | *Macromol Mater Eng* | *Polym-Korea* |
| *Iran Polym J* | *Macromol Rapid Comm* | *Polym-Plast Technol* |
| *J Appl Polym Sci* | *Macromol Symp* | *Polymer* |
| *J Inorg Organomet P* | *Macromol Theor Simul* | *Prog Polym Sci* |
| *J Macromol Sci Phys* | *Macromolecules* | *React Funct Polym* |
| *J Macromol Sci Pure* | *Mater Res Innov* | *Rubber Chem Technol* |
| *J Plast Film Sheet* | | |

**Table II**
Bi-connected component containing 49 journals about polymers

The bi-connected component can be extracted, and it then becomes clear that this group is internally structured in a densily interrelated core set (15 journals) focused on polymers and macromolecules as subjects, another set of journals which focus on the engineering side, some journals dealing with specific applications, interconnecting journals, and other more peripheral journals (Figure 8).

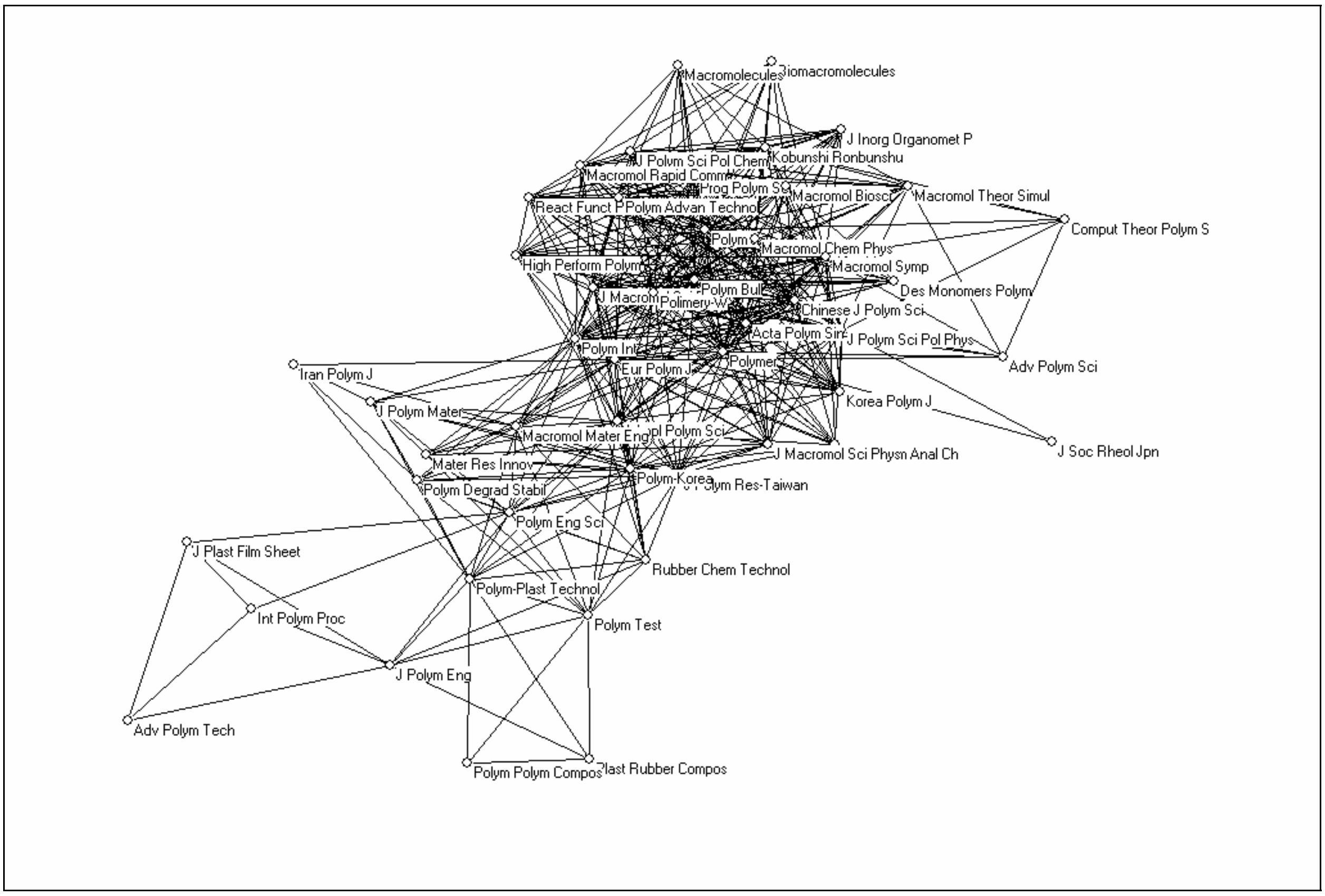

**Figure 8**

49 journals in "polymer science" form a single bi-component with an internal structure (using the Kamada & Kawai algorithm)

If we set a minimum size for the bi-connected components to ten journals, 62 bi-components are distinguished that are very homogenous and meaningful; but the groups vary in size. 2,726 journals are organized in these 62 groupings. Only seven articulation points among these larger groups can be identified. As noted, the large bi-component with 1,417 journals will be further decomposed into subcategories in the next section. Appendix One provides the categorization of all groups and subgroups.[8] Table III summarizes the statistics for the larger set of bi-components with three journals and the smaller set with a minimum of ten journals, respectively.

[8] The corresponding journal mappings are brought on-line at http://www.leydesdorff.net/jcr01 .

| 3991 journals from the Science Citation Index | *Bi-components with three or more journals* | *Bi-components with ten or more journals* |
|---|---|---|
| *Nr of bi-components* | 222 | 62 |
| *Nr of journals included* | 3447 | 2726 |
| *Nr of articulation points* | 55 | 7 |

**Table III**

Number of bi-components and articulation points given different minimum sizes

Figure 9 shows the 55 articulation points, which are almost exclusively isolates. Thus, the relations between the various clusters are maintained (at this level of correlation) by single journals. These journals have a function at a specific interface (Leydesdorff & Cozzens, 1993).

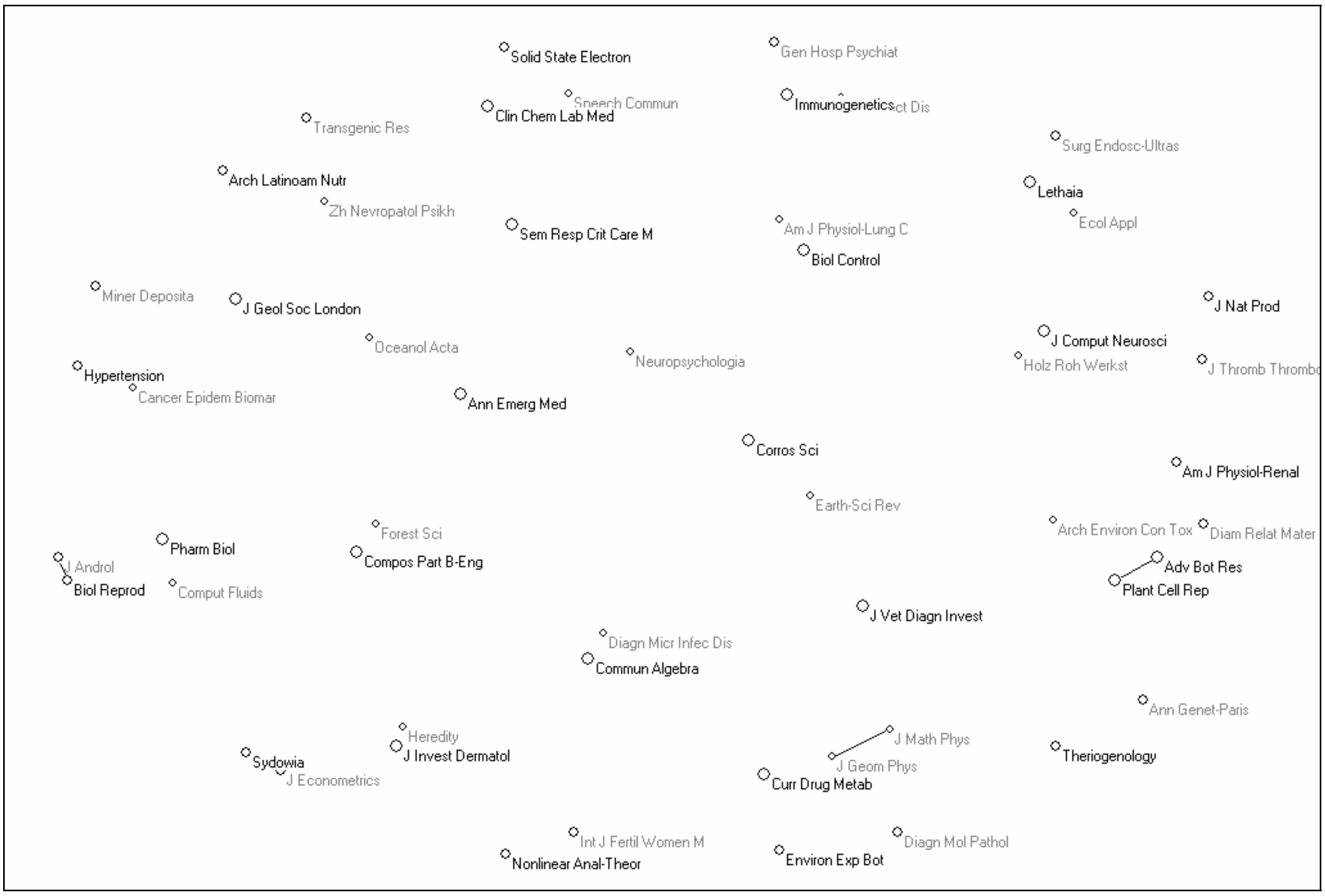

**Figure 9**

55 articulation points among 222 bi-connected components (using a 3-dimensional solution of the Fruchterman-Reingold algorithm for the visualization)

In summary, the next-order nodes are scattered over the database and decentralized. They do not indicate a next-order level. Furthermore, none of the bi-connected components can be recognized as performing specifically this function of a next-order group. However, there is one bi-connected component which is more strongly connected than the other 61, notably the large one of 1,417 journals. We turn now to the decomposition of this cluster.

### *4.4 Further decomposition of the large bi-component*

The large bi-component of 1,417 journals can further be decomposed by raising the threshold of the Pearson correlation to the level of $r \geq 0.9$. 426 of these 1,417 journals (30.1%) are then no longer connected. Thus, the analysis can be pursued using the 991 journals (69.9%) remaining in the network.

776 of these 991 journals can be organized into 51 bi-connected components. 18 of these 51 clusters have a size of ten or more journals; they contain 642 journals. These 18 clusters are listed in the Appendix and further analyzed. One of these clusters (nr. 6) still contains 177 journals and will further be decomposed at the level of $r \geq 0.95$.[9]

Only ten articulation points are distinguished among the 51 bi-components at the size level of three journals. These ten journals are listed in Table IV. Again, they seem to function at specific interfaces.

| |
|---|
| *Blood* |
| *Crit Rev Immunol* |
| *Classical Quant Grav* |
| *Phys Rev A* |
| *Differentiation* |
| *Eur J Hum Genet* |
| *Nat Genet* |
| *Environ Microbiol* |
| *J Microbiol Meth* |
| *Can Med Assoc J* |
| **Table IV**<br>Ten association points among 776 journals related at $r \geq 0.9$ |

[9] At the level of $r \geq 0.85$, 1,030 journals are included in five clusters. The largest cluster then still contains 523 journals, followed by a second one comprising 142 journals.

As noted, one of the clusters is still large (177 journals in 'molecular biology') and can be broken down into three subclusters at the level of $r \geq 0.95$. On the basis of the titles included, the three subclusters can be labeled as 'biochemistry,' 'cell biology,' and 'cancer research,' respectively.

The second large cluster is composed of 73 journals and can be considered as a cluster of chemistry journals. Figure 10 provides this map as another illustration.

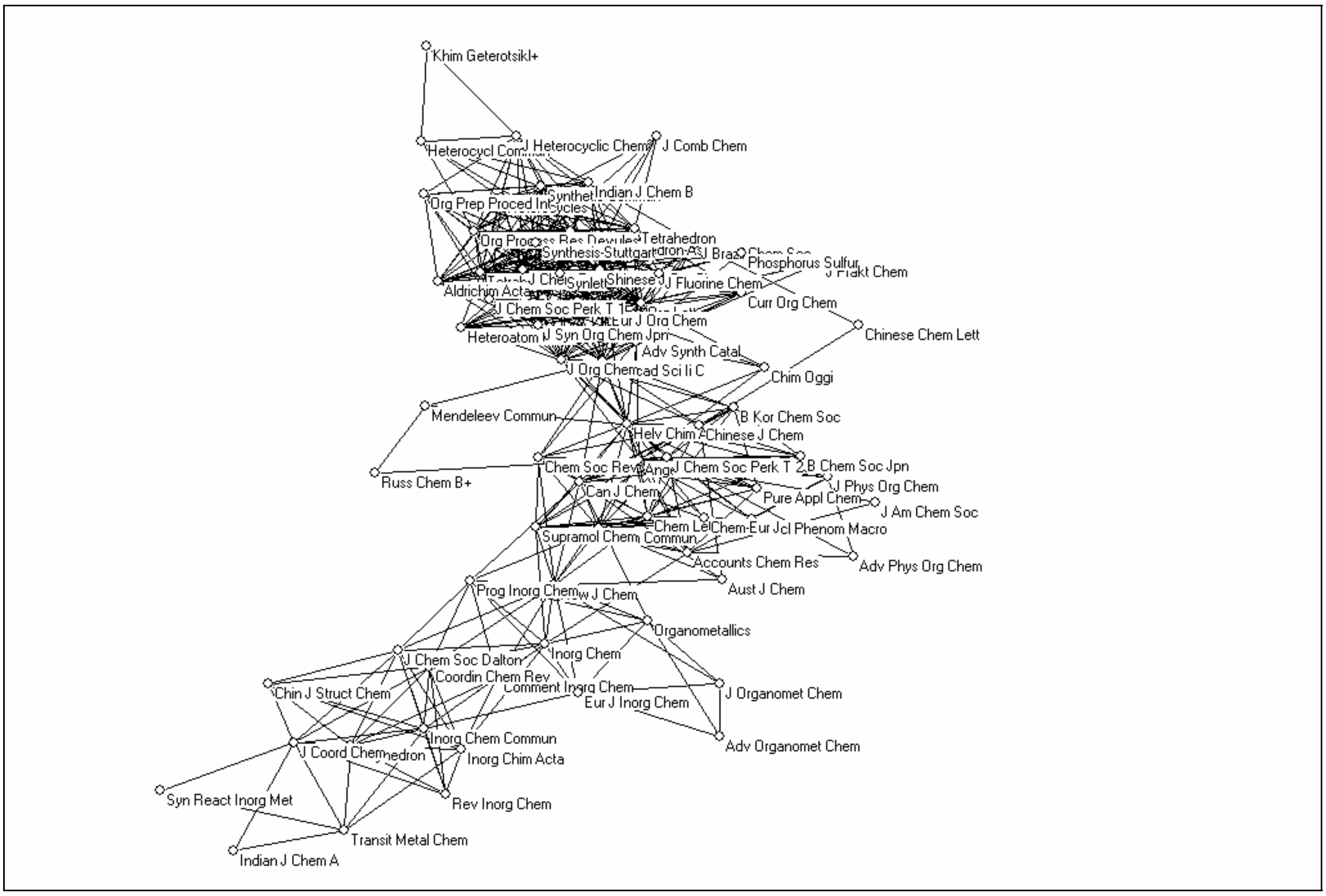

**Figure 10**

73 journals in the chemistry domain related at the level of $r \geq 0.9$ (using the Kamada & Kawai algorithm)

The map shows a grouping of inorganic chemistry journals towards the top-left corner and a grouping of organic chemistry journals towards the bottom-right corner. In between are the journals which belong to the discipline of chemistry with other designated specialties. Analytical chemistry, physical chemistry, chemical engineering, and biochemistry journals, however, are sorted into separate bi-connected components.

## 5. Conclusion

The long-standing problem in scientometrics of how to cluster the database of journals contained in the *Journal Citation Reports* of the *Science Citation Index* can be solved by using the algorithm of bi-connected components with articulation points from graph theory. A bi-connected component provides a robust definition of a cluster. The clusters are an order of magnitude larger than the fine structures made visible by using factor analysis in local environments (Leydesdorff & Cozzens, 1993; Leydesdorff, 2004). Clusters of journals can be mapped using software for the visualization. Core/periphery structures can be indicated within each of the sets.

Articulation points are relatively scarce among the larger bi-components. For example, there were only seven articulation points between the 62 bi-components with ten or more journals distinguished in the analysis of bi-components with ten or more journals. In other words, the next-order journals function at specific interfaces and do not form another layer of interaction among them. Thus, we may conclude that the journal-journal citation matrix is nearly decomposable (Simon, 1973). This indicates that the system can be considered as mainly differentiated in terms of evolutionary theorizing and therefore no longer constrained by hierarchical patterns at the level of the database of 5500+ journals (Leydesdorff, 2001, 2002b).

This conclusion raises the question of the position of so-called 'general science' journals like *Science* and *Nature* (Carpenter & Narin, 1973). These journals are not classified as a separate group by using these methods. They are part of the large group of 1,417 journals related in a single bi-component at the level of $r \geq 0.8$. Some of the specialist journals published by the *Nature* group such as *Nature Cell Biology* are visible in the Appendix as belonging to the 'molecular biology' sub-grouping of 177 journals (nr. 15.6) before its further decomposition at the level of $r \geq 0.95$. The *Proceedings of the National Academy of Sciences of the U.S.A.* also belongs to this same grouping. However, these 'general science' journals do not function as articulation points relating otherwise discrete graphs at a hierarchically next-order level. The journal literature is highly specialized and therefore specifically clustered.

**References**

Ahlgren, P., Jarneving, B. & Rousseau, R. (2003), "Requirement for a Cocitation Similarity Measure, with Special Reference to Pearson's Correlation Coefficient", *Journal of the American Society for Information Science and Technology,* Vol. 54 No. 6, pp. 550-560.

Borgatti, S. P., Everett, M. G. & Freeman, L. C. (2002), *Ucinet for Windows: Software for Social Network Analysis*. Analytic Technologies, Harvard.

Burt, R. S. (1982), *Toward a Structural Theory of Action* New York, etc.: Academic Press.

Carpenter, M. P., & Narin, F. (1973), "Clustering of Scientific Journals", *Journal of the American Society of Information Science,* Vol. 24, pp. 425-436.

Chen, C. (2003). *Mapping Scientific Frontiers: The Quest for Knowledge Visualization*. Springer, London.

Cozzens, S. E. (1985), "Using the Archive: Derek Price's Theory of Differences among the Sciences", *Scientometrics*, Vol. 7, pp. 431-441.

Doreian, P. (1986), "A Revised Measure of Standing of Journals in Stratified Networks", *Scientometrics*, Vol. 11, pp. 63-72.

Doreian, P., & Fararo, T. J. (1985), "Structural Equivalence in a Journal Network", *Journal of the American Society of Information Science,* Vol. 36, pp. 28-37.

Fruchterman, T., & Reingold, E. (1991), "Graph Drawing by Force-Directed Replacement", *Software--Practice Experience*, Vol. 21, pp. 1129-1166.

Garfield, E. (1972), "Citation Analysis as a Tool in Journal Evaluation", *Science,* Vol. 178, pp. 471-479.

Garfield, E. (1979), *Citation Indexing: Its Theory and Application in Science, Technology, and Humanities*, John Wiley, New York.

Garfield, E., & Sher, I. H. (1963), "New Factors in the Evaluation of Scientific Literature through Citation Indexing", *Am. Doc.,* Vol. 14, pp. 191.

Glänzel, W., & Schubert, A. (2003). "A New Classification Scheme of Science Fields and Subfields Designed for Scientometric Evaluation Purposes," *Scientometrics*, Vol. 56, No. 3, pp. 357-367.

Hirst, G. (1978), "Discipline Impact Factors: A Method for Determining Core Journal Lists", *Journal of the American Society for Information Science,* Vol. 29, pp. 171-172.

Kamada, T., & Kawai, S. (1989), "An Algorithm for Drawing General Undirected Graphs", *Information Processing Letters*, Vol. 31 No. 1, pp. 7-15.

Knaster, B. & Kuratowski, C. (1921), "Sur les ensembles connexes", *Fund. Math.* Vol. 2, pp. 206-255.

Leydesdorff, L. (1986), "The Development of Frames of References", *Scientometrics,* Vol. 9, pp. 103-125.

Leydesdorff, L. (1987), "Various Methods for the Mapping of Science", *Scientometrics,* Vol. 11, pp. 291-320.

Leydesdorff, L. (2001), *A Sociological Theory of Communication: The Self-Organization of the Knowledge-Based Society*, Universal Publishers, Parkland, FL; <http://www.upublish.com/books/leydesdorff.htm >.

Leydesdorff, L. (2002a), "Indicators of Structural Change in the Dynamics of Science: Entropy Statistics of the *SCI Journal Citation Reports*", *Scientometrics*, Vol. 53, No. 1, pp. 131-159

Leydesdorff, L. (2002b), "Dynamic and Evolutionary Updates of Classificatory Schemes in Scientific Journal Structures", *Journal of the American Society for Information Science and Technology*, Vol. 53 No. 12, pp. 987-994.

Leydesdorff, L. (2003), "Can Networks of Journal-Journal Citations Be Used as Indicators of Change in the Social Sciences?" *Journal of Documentation,* Vol. 59 No. 1, pp. 84-104.

Leydesdorff, L. (2004). "Top-Down Decomposition of the Journal Citation Report of the Social Science Citation Index: Graph- and Factor-Analytical Approaches", *Scientometrics*, in print.

Leydesdorff, L., & Cozzens, S. E. (1993), "The Delineation of Specialties in Terms of Journals Using the Dynamic Journal Set of the Science Citation Index", *Scientometrics,* Vol. 26, pp. 133-154.

Leydesdorff, L., Cozzens, S. E. & Van den Besselaar, P. (1994), "Tracking Areas of Strategic Importance Using Scientometric Journal Mappings." *Research Policy*, Vol. 23, pp. 217-229.

Leydesdorff, L., & Gauthier, É. (1996), "The Evaluation of National Performance in Selected Priority Areas Using Scientometric Methods", *Research Policy*, Vol. 25, pp. 431-450.

Leydesdorff, L., & Jin, B. (in preparation), "Mapping the *Chinese Science Citation Database* in terms of aggregated journal-journal citation relations".

Leydesdorff, L., & Zaal, R.. (1988), "Co-Words and Citations. Relations between Document Sets and Environments," in Egghe, L., and Rousseau, R. (Eds.), *Informetrics 87/88.* Elsevier, Amsterdam, pp. 105-119.

Moed, H. F. (1989), "Bibliometric Measurement of Research Performance and Price's Theory of Differences among the Sciences", *Scientometrics,* Vol. 15 Nos. 5-6, pp. 473-483.

Moed, H. F., Burger, W. J. M., Frankfort, J. G. & Van Raan, A. F. J. (1985), "The Use of Bibliometric Data for the Measurement of University Research Performance", *Research Policy,* Vol. 14, pp. 131-149.

Moody, J. & White, D. R. (2003), "Structural Cohesion and Embeddedness: A Hierarchical Concept of Social Groups", *American Sociological Review,* Vol. 68 No. 1, pp. 103-127.

Mrvar, A., & Bagatelj, V. (n.d.), "Network Analysis Using Pajek"; at http://vlado.fmf.uni-lj.si/pub/networks/pajek/doc/pajekman.htm . Retrieved December 9, 2003

Narin, F. (1976), *Evaluative Bibliometrics: The Use of Publication and Citation Analysis in the Evaluation of Scientific Activity*. National Science Foundation, Washington, DC.

Narin, F., Carpenter, M., & Berlt, N. C. (1972), "Interrelationships of Scientific Journals", *Journal of the American Society for Information Science*, Vol. 23, pp. 323-331.

Noma, E. (1982), "An Improved Method for Analyzing Square Scientometric Transaction Matrices", *Scientometrics,* Vol. 4, pp. 297-316.

Owen-Smith, J., Riccaboni, M., Pammolli, F., & Powell, W. W. (2002), "A Comparison of U.S. And European University-Industry Relations in the Life Sciences", *Management Science*, Vol. 48, No. 1, pp. 24-43.

Price, D. J. de Solla (1965), "Networks of Scientific Papers", *Science,* Vol. 149, pp. 510-515.

Price, D. J. de Solla (1970), "Citation Measures of Hard Science, Soft Science, Technology, and Nonscience", in Nelson, C. E. & Pollock, D. K. (Eds.), *Communication among Scientists and Engineers.* Heath, Lexington, MA, pp. 3-22.

Price, D. J. de Solla (1981), "The Analysis of Square Matrices of Scientometric Transactions", *Scientometrics,* Vol. 3, pp. 55-63.

Scott, J. (1991). *Social Network Analysis*. Sage, London, etc.

Simon, H. A. (1973), "The Organization of Complex Systems", in Pattee, H. H. (Ed.), *Hierarchy Theory: The Challenge of Complex Systems*. George Braziller Inc., New York, pp. 1-27.

Small, H. (1999), "Visualizing Science by Citation Mapping", *Journal of the American Society for Information Science*, Vol. 50, No. 9, pp. 799-813.

Small, H., & Griffith, B. (1974), "The Structure of the Scientific Literature I", *Science Studies,* Vol. 4, pp. 17-40.

Small, H., & Sweeney, E. (1985), "Clustering the Science Citation Index Using Co-Citations I. A Comparison of Methods", *Scientometrics,* Vol. 7, pp. 391-409.

Small, H., Sweeney, E. & Greenlee, E. (1985), "Clustering the Science Citation Index Using Co-Citations II. Mapping Science", *Scientometrics,* Vol. 8, pp. 321-340.

Studer, K. E., & Chubin, D. E. (1980), *The Cancer Mission. Social Contexts of Biomedical Research*. Sage, Beverly Hills, etc.

Tijssen, R., De Leeuw, J., & Van Raan, A. F. J. (1987), "Quasi-Correspondence Analysis on Square Scientometric Transaction Matrices", *Scientometrics,* Vol. 11, pp. 347-361.

White, H. D. (2003), "Author Cocitation Analysis and Pearson's *r*", *Journal of the American Society for Information Science and Technology*, Vol. 54, No. 13, pp. 1250-1259.